\documentclass[aps,prl,showpacs,amsmath,amssymb,amsfonts,lengthcheck,twocolumn,longbibliography,superscriptaddress]{revtex4-2}
\usepackage{epsfig,graphicx,graphics,amsmath,amssymb,float}
\usepackage[T1]{fontenc}
\usepackage[latin9,utf8]{inputenc}
\usepackage{xcolor}
\usepackage{amscd}
\usepackage{bm}
\usepackage{bbold}
\usepackage{psfrag}
\usepackage{mathrsfs}
\usepackage{float}
\usepackage{svg}

\usepackage{graphicx}
\usepackage{braket}
\usepackage[colorlinks=true, allcolors=blue]{hyperref}

\usepackage{dsfont}

\begin{document}

%\title{Entangleability is an energetic control resource}
%\title{The energetic advantage of monolithic quantum gates} - maybe this would be more appropriate for a follow-on paper 
%\title{A universal quantum computer requires no energy}
%\title{Energetic advantage of entangleability in Hamiltonian quantum gates}
\title{Hamiltonian quantum gates -- energetic advantage from \emph{entangleability}}
\author{Josey Stevens}
\email{Josey.Stevens@jhuapl.edu}
\affiliation{Johns Hopkins University Applied Physics Laboratory, Laurel, Maryland 20723, USA}
\affiliation{Department of Physics, University of Maryland, Baltimore County, Baltimore, MD 21250, USA}
\affiliation{Quantum Science Institute, University of Maryland, Baltimore County, Baltimore, MD 21250, USA}

\author{Sebastian Deffner}
\email{deffner@umbc.edu}
\affiliation{Department of Physics, University of Maryland, Baltimore County, Baltimore, MD 21250, USA}
\affiliation{Quantum Science Institute, University of Maryland, Baltimore County, Baltimore, MD 21250, USA}
\affiliation{National Quantum Laboratory, College Park, MD 20740, USA}
\date{\today}

\begin{abstract}
Hamiltonian quantum gates controlled by classical electromagnetic fields form the basis of any realistic model of quantum computers.  
In this letter, we derive a lower bound on the field energy required to implement such gates and relate this energy to the expected gate error. 
We study the \emph{entangleability} (ability to entangle qubits) of Hamiltonians and highlight how this feature of quantum gates can provide a means for more energetically efficient computation.  
Ultimately, we show that a universal quantum computer can be realized with vanishingly low energetic requirements but at the expense of arbitrarily large complexity.  
\end{abstract}

\maketitle

%\section{Introduction}

For certain problems, quantum algorithms are generally expected to exhibit significant computational advantage over classical computing \cite{Nielsen_Chuang_2010}.
This ``quantum advantage'' can be quantified by counting the number of quantum gates required to implement a specific algorithm, and compares this number to the necessary single gate operations in the best known classical algorithms. Even once the overhead of quantum error correction \cite{gottesman2002introduction} is taken into account, quantum algorithms are still expected to retain their significant advantage, and thus quantum computers are expected to revolutionize computation into the future \cite{Sanders2017}. 

This ``gate counting'' approach to quantum advantage is perfectly adequate to assess many resource requirements of quantum computers.
For example, if the requirements for a set of universal quantum gates can be estimated, the quantitative assessment of a fault-tolerant quantum algorithm follows directly as some scaled factor, which depends on the error sources and error correction scheme \cite{10.1126/science.279.5349.342, doi:10.1137/S0097539799359385}. However, any such estimate of the total resource requirements can only be determined in relation to the fundamental ``cost'' of a single gate operation.

Thus, it appears obvious that robust estimates for the energetic, thermodynamics, as well as computational resources for single gates is necessary to meaningfully assess the performance of quantum computers. In fact, it has been recently recognized that an ``quantum energy initiative'' is urgently required so as not to stall the development of quantum computing \cite{Auffeves2022PRXQ}. Interestingly, energetic considerations were initially secondary concerns in computer designs, but as semiconductor devices advanced and semiconductor physics were pushed, both Moore's law and Denard scaling broke down \cite{10.1109/n-ssc.2006.4785860,10.1145/2024723.2000108}.
Recent evaluations \cite{malmodin2024ict} have shown that information and communication technology is consuming nearly four percent of global electric power consumed, and this demand is expected to grow in the future \cite{IEA2024, IEA2025}. Consequently, evaluating the energetic demands of quantum computing could be vitally important to predict future economic, energy and technology trends.

Previous characterizations of the energy requirements for quantum technologies \cite{Auffeves2022PRXQ,Deffner2019book} include, e.g., extensions of Zurek's bound on dissipated energy \cite{Zurek1989,zurek1989algorithmic,kolchinsky2023generalized,zhu2024} and quantum speed limits \cite{10.1007/s11128-016-1405-x, 10.22331/q-2019-08-05-168,Deffner_2017} that characterize the energy in the control field required for computation.
These bounds, however, require precise knowledge of the quantum state before and after a computation, information which is unavailable in the middle of a real computation.
Alternatively, extensions of Landauer's bound \cite{Bedingham_2016,Goold_NEQ_Landaurer,PhysRevLett.124.240601,PhysRevLett.128.010602,10.1088/1361-6633/add6b3,PhysRevA.107.042202,PhysRevA.96.042109,PhysRevLett.128.010602,reeb2014improved,PhysRevE.83.030102}
provide estimates of the dissipated energy, but only consider the width of a quantum algorithm without any consideration of gate counts.
Recent approaches to consider the energetic requirements of quantum computation based on Schatten-$P$ norms have been explored in Refs.~\cite{PhysRevA.94.042132,Aifer_2022,Deffner_2021}; yet, these approaches are either inconsistent when including a non-interacting universe or weak for complex quantum operations.

For our purposes, we are motivated by Ref.~\cite{PhysRevLett.89.217901}, which proposed a lower bound for the control energy required to implement the conditional sign-flip gate.  In the present analysis, we generalize the approach of Ref.~\cite{PhysRevLett.89.217901} into a formal method for characterizing the energy requirements for any arbitrary quantum computation. To this end, we make only minimal assumptions, namely that external control is performed via a classical electromagnetic field. We then show show that, generally, the energy requirements are inversely proportional to the square of the gate errors introduced by fluctuations of this field.
We consider both single and multi-qubit gates in a single framework allowing for flexibility in assessing unitary quantum algorithms via any gate decomposition.  

% We begin in Section \ref{sec Framework} by defining our theoretical framework.
% In Section \ref{Gate Errors} we provide limitations on the gate errors arising from the fluctuations of the control field.
% Lower bounds on the required energy are derived in Section \ref{Sec: Required Energy}.
% And we conclude with an assessment of entangling Hamiltonians and universal quantum computation in Section \ref{Sec entanglement}.

\paragraph{Coherent Hamiltonian quantum control}\label{sec Framework}

Current quantum computing designs utilize externally driven classical fields rather than carefully prepared quantum interactions.
With this point in mind, we consider our quantum computer as $N$-qubits coupled to a multimode coherent electromagnetic field with standard definition $\ket{\psi(t)} = \prod_k \ket{\alpha_k(t)}$ and the requirement that each coherent mode obeys $a_k\ket{\alpha_k(t)}=\alpha_k\ket{\alpha_k(t)}$ with $a_k$ and $\alpha_k(t)$ as the lowering operator and time-dependent amplitude of the $k$th mode, respectively.
Multimode coherent states describe fields created by classical electric currents and strong laser pulses, which are utilized, for instance, to control superconducting circuits \cite{Li_2021,li2019manipulation,raftery2017direct,bardin2021microwaves} and trapped-ion qubits \cite{pogorelov2021compact,debnath2016demonstration}, respectively.

In the interaction picture, the coupling Hamiltonian between the field and qubits can be written as,
\begin{equation} \label{eq hamiltonian}
    H_{\text{int}} = \hbar \sum_{i} \sum_k g_{k,i} \left( a_k e^{-i \omega_k t} + a^\dag _k e^{i \omega_k t} \right) V_i,
\end{equation}
where we have expanded the Hamiltonian as a sum over commuting operators, $V_i$, which act on the $N$-qubit Hilbert space. The coupling strength between the operators and field mode with frequency $\omega_k$ is given by $g_{k,i}$.

We now implement a unitary quantum gate $U_g$ in time $\tau$ by enforcing that $U_G$ be equal to the time evolution operator evaluated at time $\tau$,
\begin{equation}  \label{UgDef}
    U_G = \mathcal{T}_>\exp{\left(-\frac{i}{\hbar}\int_0^\tau  H_{\text{int}}(t) \, dt\right)},
\end{equation}
from which we calculate (on the principle branch),
\begin{equation} \label{LogUgDef}
    \text{ln}(U_G) = -i\sum_{i,k}  g_{k,i} \int_0^\tau   \left( a_k e^{-i \omega_k t} + a^\dag _k e^{i \omega_k t} \right) V_i \, dt.
\end{equation}
Note that the matrix logarithm is not unique, which will become instrumental in the following discussion.

\paragraph{Tensor product decomposition}

A straightforward approach is the eigendecomposition of $\text{ln}(U_G)$. However, this choice leads to an inconsistent measure of the required energy when including non-interacting qubits in our control Hamiltonian due to the repeated eigenvalues which arise when representing the Hamiltonian in the extended basis.  Instead we use a tensor-product decomposition \cite{10.1002/sapm192761164}, which expands a generically multipartite operator into a sum over operators,$C_{i,j}$, acting in parallel on sub-systems,
\begin{equation} \label{eq tensor product decomposition}
    A  = \sum_i \lambda_i \bigotimes_{j=1}^N C_{i,j},
\end{equation}
where we require the operators to commute,
\begin{equation}
\left[\bigotimes_{j=1}^N C_{i,j},\bigotimes_{j=1}^N C_{k,j}\right] = 0\,.
\end{equation} 
A useful (though not unique) basis for such a decomposition is given by the multi-qubit Pauli matrices. From Eqs.~\eqref{LogUgDef} and \eqref{eq tensor product decomposition} we then immediately identify the interaction operators, $V_i = \bigotimes_{j=1}^N C_{i,j}$, and corresponding coefficients,
\begin{equation} \label{Eq: Field Term Relationships}
    \lambda_i = -\int_0^\tau \sum_k g_{k,i}   \left( a_k e^{-i \omega_k t} + a^\dag _k e^{i \omega_k t} \right)\, dt ,
\end{equation}
which must hold for all terms of our decomposition.

If we take the view that $\lambda_i$ is an observable of the coherent field, we immediately conclude that it is subject to a distribution of measurement outcomes, i.e., effects on the quantum gate, due to fluctuations in the quantum field.
To characterize this distribution we compute the variance as,
\begin{equation} \label{Eq: ErrorTerm}
     \text{var}(\lambda_i) = \sum_k \left|\int_o^\tau g_{k,i}e^{-i \omega_k t} \,dt \right|^2 \sim \epsilon_i^2,
\end{equation} 
 where we have identified the typical error in $\lambda_i$ as $\epsilon_i$ which is nearly always the order of magnitude of the standard deviation in accordance with Chebychev's inequality \cite{Mitzenmacher2017-mc}. A derivation of  Eq.~\eqref{Eq: ErrorTerm} is provided in the End Matter and we note that a version of this result was previously provided by Gea-Banecloche \cite{PhysRevLett.89.217901} with a different association to the gate error. 
 In the next section, we will explore how the error on individual terms via $\lambda_i$ contributes to our overall gate error.

\paragraph{Sub-linearity of gate errors}\label{Gate Errors}

While Eq.~\eqref{Eq: Field Term Relationships} provides the necessary conditions to implement $\text{U}_G$ exactly, in practice thermal and quantum noise result in an imperfectly realized gate, $\text{U}_{G^{'}}$, defined by
\begin{equation} \label{Ugprime}
    \text{U}_{G^{'}} = \exp{\left(i\sum_i (\lambda_i + \epsilon_i)V_i\right)}.
\end{equation}
If our gate had been implemented without errors, we would have $\text{U}_{G^{'}}\text{U}_G^{\dag} = \mathds{1} $. For imperfect gates, we expect $\text{U}_{G^{'}}\text{U}_G^{\dag}$ to differ from identity by an operator with a small norm.
We identify the norm of this operator as the error of the gate, $\epsilon$, which can be calculated as,
\begin{equation} \label{Error Definition}
    \epsilon = \left\Vert \mathds{1} - \text{U}_{G^{'}} \text{U}_G^{\dag}\right\Vert_\infty,
\end{equation}
where $\left\Vert \cdot \right\Vert_\infty$ is the operator norm.  
We observe that $\text{U}_{G^{'}} \text{U}_G^{\dag}$  resembles the time evolution operators in the Loschmidt echo $\mathcal{L}(\tau) = \left|\bra{\psi} e^{i H_2 \tau/\hbar}e^{-i H_1 \tau/\hbar}\ket{\psi}\right|^2$ \cite{PhysRevA.30.1610} which characterizes the overlap of the forward and backward time evolution of two dissimilar Hamiltonians for a given input state and note that for pure input states the Loschmidt echo is identical to the fidelity of two evolved states.
It can be shown that $\epsilon \ge \left|1-\sqrt{\mathcal{L}(\tau)}\right|$ for all possible input states, a proof of this statement is provided in the End Matter.
Thus, we immediately understand that $\epsilon$ correctly characterizes the error of the implemented gate without reference to any specific input state.

We can now use Eqs.~\eqref{Eq: ErrorTerm} and \eqref{Ugprime} together with the commutation properties of our chosen basis, $V_i$, to express $\epsilon$ explicitly as,
\begin{equation}
\begin{aligned} \label{eq epExplicit}
    \epsilon &= \left\Vert \mathds{1} -\prod_i\left( \sum_{k=0}^\infty  \frac{i^k \epsilon_i^k}{k!} V_i^k  \right)\right\Vert_\infty\\
     &= \left\Vert \mathds{1} -\prod_i\left( i \sin{(\epsilon_i)} V_i + \cos{(\epsilon_i)} \mathds{1} \right)\right\Vert_\infty,
 \end{aligned}
 \end{equation}
where in the second step we have also assumed that our operators, $V_i$, are involutory , i.e., they are their own inverse, a condition satisfied by the multi-qubit Pauli matrices, among others \cite{higham2008functions}.
By retaining only leading orders of $\epsilon_i$ and utilizing the subadditivity of the matrix norm \cite{Horn1985-ro}, we immediately have,

\begin{equation}\label{eq error bound}
    \epsilon \le \sum_i \left|\epsilon_i\right|,
\end{equation}
which demonstrates that our gate error is upper bounded by the sum of the independent error terms.  
Combined with the fact that Eq.~\eqref{Error Definition} upper bounds the realized computational error, we immediately see that the realized gate error, characterized by the fidelity between the desired and achieved quantum state, is upper-bounded by the sum of the independent error terms realized in the driving field.

%\subsection{Matrix Logarithm and the Problem Of Choice}

\paragraph{Required energy} \label{Sec: Required Energy}

Having established our gate decomposition and error relations, we can now turn toward our main objective, namely determining a rigorous bound for the energetic requirements of implementing a specific gate.
Our approach will be to determine the energy required to implement a single control coefficient $\lambda_i$, and then, once the fundamental relationships are established, to bootstrap the energy relationships into a series of bounds, depending on how the fields couple to the qubit system.

To this end, we start by bounding the square of Eq.~\eqref{Eq: Field Term Relationships} by,
\begin{equation}
\begin{aligned}
    \frac{\left|\lambda_i\right|^2}{4} &\le \left|\int_0^\tau  \sum_k g_{k,i}    a_k e^{-i \omega_k t}\,dt  \right|^2\\
    &\le \left(\sum_k  \left|a_k\right|^2\right)  \left(\sum_k \left|\int_0^\tau  g_{k,i}    e^{-i \omega_k t} \,dt \right|^2\right),
\end{aligned}
\end{equation}
where in the last step we utilize the Cauchy-Schwarz inequality to separate the field amplitudes from the coupling and time dependent terms of the field.
We can immediately identify from Eq.~\eqref{Eq: ErrorTerm} that the right most summation is simply $\epsilon_i^2$, and  with some algebra, which the aim of will become clear in the next step, arrive at,
\begin{equation} 
         \frac{\hbar \left|\lambda_i\right|^2}{4 \epsilon_i^2} \le \hbar \sum_k \frac{\omega_k}{\omega_k} \left| a_k \right|^2.
\end{equation}
By utilizing the fact that our field frequencies are ordered in magnitude we can further simplify this result to,
\begin{equation} \label{eq energy derrivation 1}
         \frac{\hbar \omega_0 \left|\lambda_i\right|^2}{4 \epsilon_i^2} \le \hbar \sum_k \omega_k \left| a_k \right|^2.
\end{equation}

Recalling that the average field energy \cite{Loudon2000-qd} is given by,
\begin{equation} \label{eq field energy def}
    \braket{E_i} = \hbar \sum_k \omega_k \left| a_k \right|^2,
\end{equation}
and substituting this definition into Eq.~\eqref{eq energy derrivation 1} we find,
\begin{equation} \label{Eq independent energy bound}
    \braket{E_i} \ge \frac{\hbar \omega_0 \left| \lambda_i\right|^2}{4 \epsilon_i^2}.
\end{equation}
This is our first step toward a general bound, i.e., a bound on the energy needed to implement a single mode of a gate.

If our chosen gate and decomposition requires more than one field coupling, i.e., some $\lambda_{i>1} \ne0$, we must include the energetic and error terms of these couplings together.
First, consider the case where $\lambda_i$ are controlled by independent fields with no coupling terms $g_i,k$ being shared between parallel acting operators, i.e.,$g_{k,i}*g_{k,j}\propto\delta_{i,j}$.  
In this case, the energetic requirement from Eq.~\eqref{Eq independent energy bound} sums over parallel operators and we obtain,
\begin{equation} \label{Eq Ind Energy Bound combined}
\begin{aligned}
    \braket{E} &\ge \frac{\hbar}{4}\sum_i \frac{\omega_i \left|\lambda_i\right|^2}{\epsilon_i^2} \ge \frac{\hbar \omega_0}{4 \epsilon^2}\sum_i \left|\lambda_i\right|^2,
\end{aligned}
\end{equation}
where in the last step we have realized a weaker, but simpler, bound by utilizing the fact that $\epsilon \ge \epsilon_i$ and that our field frequencies are ordered.

If instead of independent control fields, the fields couple to more than one term in our decomposed Hamiltonian, the same field energies are used to implement multiple control modes.
In the case of shared couplings, we see that the control can be more energetically efficient than Eq.~\eqref{Eq Ind Energy Bound combined}.
An exact form of the bound in this case depends on the nature of the coupling, and for this reason we will consider only the most extreme case where all tensor-decomposed modes couple to the field modes equally.
In this case, the bounds of Eq.~\eqref{Eq independent energy bound} hold for each term independently and the total energy bound is given by,
\begin{equation} \label{Eq shared energy bound}
    \braket{E} \ge \frac{\hbar \omega_0 }{4 \epsilon^2} \max_i(\left| \lambda_i\right|^2),
\end{equation}
where we have again utilized the fact that $\epsilon \ge \epsilon_i$.

Inequalities~\eqref{Eq Ind Energy Bound combined} and \eqref{Eq shared energy bound} constitute the first main result, a bound on the energy that must be contained in a coherent electromagnetic field to implement an arbitrary quantum gate via Hamiltonian control.
This result depends only on the decomposition of the desired gate, the lowest frequency field mode of the field, and the error resulting from this control.
Except for the error entering as an order of magnitude, the bound tight when only a single-field mode is utilized to drive the system.  

By examining bounds \eqref{Eq Ind Energy Bound combined} and \eqref{Eq shared energy bound} we immediately see that, even without external sources of gate error, realizing increasingly accurate quantum gates requires increasingly large amounts of energy.
More specifically, each order of magnitude of decrease in error requires two orders of magnitude increase in the energy of the control fields.

\paragraph{Example: energetic cost of single qubit gates}

As a first example of this capability, we will calculate the multiset of $\lambda_i$, as expressed in the Pauli basis for many single qubit gates gates.
For the identity operator we have $\lambda_{\mathds{1} }=  \{0\}$ while for the Pauli gates themselves, we have: $\lambda_\text{X} = \lambda_Y = \lambda_Z =  \{-\frac{\pi}{2},\frac{\pi}{2} \}$.
The Hadamard gate, $H = \left(\sigma_x + \sigma_z \right)/\sqrt{2}$, and rotations about the x, y, and z-axis of the Bloch sphere have coefficient multisets given by: $\lambda_{H} =  \{\frac{\pi}{2},-\frac{\pi}{2} \}$ and $\lambda_{R_\text{X}(\theta)} = \lambda_{R_Y(\theta)} = \lambda_{R_Z(\theta)} =  \{-\frac{\theta}{2}\}$.
The Hadamard gate is the only gate presented here that requires decomposition into operators other than Pauli matrices.

By observing the coefficient sets for the single qubit gates, we see that, as expected, gates which (might) provide large changes to states require relatively large amounts of energy relative to gates which do not change any input state appreciably.

% \subsection{Energetic Requirements For A Fixed Quantum Computer Design}
% Until now, we have assumed that all the physical terms in \eqref{eq hamiltonian} were free parameters, as they would be if sitting down and designing a quantum computer without constraints.
% In contrast, designers of real quantum computers are subject to numerous design constraints.
% For example, once the basic design of the quantum computer is chosen, the coupling constants are more or less fixed and the only way to drive down errors is to utilize larger field energies.
% In this section we will derive energetic bounds assuming a design and fixed coupling constants $g_{k,i}$.  

\paragraph{Energetic benefits of entangleability} \label{Sec entanglement}

%Beyond the one-qubit gates are the multipartite gates, due to its straightforward nature, we only include the swap gate here and explore other multipartite gates shortly, where we will explore the value of \emph{entangleability}.  
%The SWAP gate has coefficients of $\lambda_{SWAP }=  \{2\cdot-\frac{\pi}{4},2\cdot\frac{\pi}{4} \}$, where we have introduced the notation $a\cdot y$ to represent $a$ repetitions of the value $y$ in a multiset.

\begin{figure}
    
    \centering
    \includegraphics[width=.48 \textwidth]{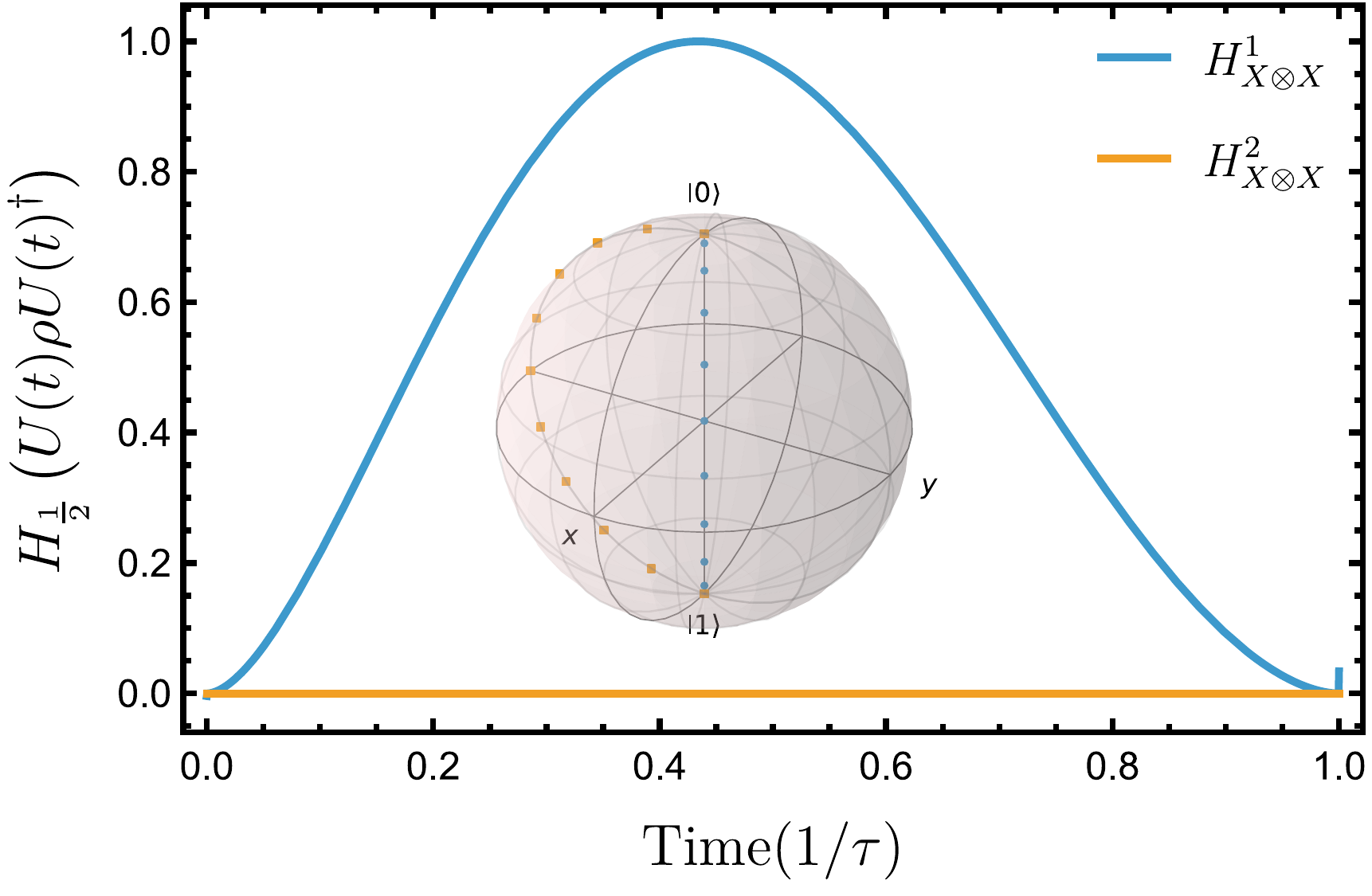}
    \caption{The Renyi-$1/2$ entropy over the duration of the driving protocol implemented by Hamiltonians  \eqref{eq Hxx1} (blue) and \eqref{eq Hxx2} (orange).  
    The initial state is prepared such that both qubits are in the zero state, i.e.,  $\rho = \ket{0 0}\bra{0 0}$ and are driven by electromagnetic fields consisting of a single mode with $\omega_0 = \frac{1}{\tau}$.  
    On the inset we show the path on the Bloch sphere of the first qubit taken at equal timesteps of  $\tau/10$.}
    \label{Fig: Entanglement}
\end{figure}

As alluded to above, the Hamiltonian satisfying Eq.~\eqref{UgDef} is not unique due to the multiple branches of the matrix logarithm.
In many cases, it suffices to discuss the principal branch of the logarithm, which returns a matrix whose eigenvalues are all imaginary with magnitude less than $\pi$ and in this way, the principal logarithm provides an efficient Hamiltonian to implement a desired gate.
Many of the non-principle branches simply result in superfluous rotations around the Bloch sphere. 

For example, from a Hamiltonian that performs a NOT gate, $U_\text{X} = \text{ln}(\text{X})$, we can generate another gate simply by multiplying the Hamiltonian by an odd integer $U^k_\text{X} = (2k + 1)U_\text{X}$.
This new Hamiltonian makes $k$ full rotations around the Bloch sphere before finally settling to a flip about the $x/y$-plane.

When considering multi-qubit gates the different branches can capture more physically rich behaviors in the computational basis.
To highlight these behaviors, we will explore the simultaneous operation of a NOT gate on the first two qubits, i.e. $\text{X} \otimes \text{X}$ via two different Hamiltonians, $H_{\text{X}\otimes \text{X}}^1$ and $H_{\text{X}\otimes \text{X}}^2$ with the following definitions:
\begin{equation} \label{eq Hxx1}
    H_{\text{X}\otimes \text{X}}^1 = \text{ln} (\text{X}\otimes \text{X}) ,
\end{equation}
and,
\begin{equation} \label{eq Hxx2}
    H_{\text{X}\otimes \text{X}}^2 = H_\text{X} \otimes \mathds{1} +  \mathds{1} \otimes H_\text{X},
\end{equation}
with $H_{\text{X}} = \text{ln} (\text{X})$.

These two distinct decompositions have the following multisets when expanded in terms of the multiqubit Pauli matrices $\lambda^1_{\text{X}\otimes \text{X}} = \{\frac{\pi}{2},\frac{\pi}{2}\}$ and $\lambda^2_{\text{X}\otimes \text{X}} = \{\pi,\frac{\pi}{2},\frac{\pi}{2}\}$, respectively.

Inspecting Eq.~\eqref{Eq shared energy bound}, we observe that $H_{\text{X}\otimes \text{X}}^1$ can be implemented for one-quarter of the energy required to implement $H_{\text{X}\otimes \text{X}}^2$, assuming that both are implemented with identical sets of driving frequencies.  
Importantly, these decompositions have different affects on the entanglement properties of the process, that is, $H_{\text{X}\otimes \text{X}}^2$ leaves the qubits unentangled throughout the entirety of evolution, while $H_{\text{X}\otimes \text{X}}^1$ entangles the qubits at all intermediate times $0\le t \le \tau$ even though at $t = 0$ and $t =  \tau$ the gates are left unentangled.
This effect can be seen in Fig.~\ref{Fig: Entanglement}, where we have implemented Hamiltonians in the specific case where that all fields have frequency $\omega_k = \frac{1}{\tau}$, and explore the paths traversed in the Bloch sphere as the systems evolve under each Hamiltonian.  
We also characterize the generated entanglement with the Renyi-$1/2$ entropy \cite{Horodecki2009RMP}, $H_\alpha (\rho) = \frac{1}{1-\alpha} \text{ln}\left( \rho^\alpha\right)$  and $\alpha=1/2$, as the systems evolve along these paths.
In this way, by utilizing an entangling Hamiltonian, the cost per qubit of implementing multi-qubit gates can be reduced.

Furthermore, if we observe the fact that $H_\text{X}$ has coefficients $\lambda_\text{X} = \{\frac{\pi}{2},\frac{\pi}{2}\}$ and comparing to $\lambda^1_{\text{X}\otimes \text{X}} = \{\frac{\pi}{2},\frac{\pi}{2}\}$ we have a noteworthy result: A NOT gate can be operated on two qubits simultaneously for the same energy required to implement a single NOT gate on one qubit.
However, if we wish to perform the two simultaneous NOT gates without entanglement (which would be left as a residual with any gate errors), we can at most perform two NOT gates with double the energetic cost of a single NOT gate.

To further highlight how entanglement can affect the energetic requirements for a gate, we will explore optimal decompositions of controlled-NOT gates.
We specify generic $\text{C}^n\text{X}$ gates, for example: $n=0$, $n=1$, and $n=2$ are the NOT, controlled-NOT, and Toffoli gates, respectively \cite{10.1007/3-540-10003-2_104}.
By expanding in terms of multi-qubit Pauli matrices, we obtain the decompositions of $\lambda_{\text{C}^1\text{X}} = \{\frac{\pi}{4},\frac{\pi}{4},-\frac{\pi}{4},-\frac{\pi}{4}\}$ and $\lambda_{\text{C}^2\text{X}} = \{4\cdot\frac{\pi}{8},4\cdot-\frac{\pi}{8}\}$. 
Generally, we find $\lambda_{\text{C}^n\text{X}} = \{2^n \cdot\frac{\pi}{2^{n+1}},2^n \cdot-\frac{\pi}{2^{n+1}}\}$.
From this decomposition, we see that a NOT gate can be implemented with arbitrarily low energy, provided that it is driven in contact with a large number of ancilla qubits, all prepared in the $\ket{1}$ state.

In this case, the ancilla qubits can be understood as performing not only the logical control of the NOT gate but also the physical control, with the external control field orchestrating the interactions.  
If the field modes are shared as in Eq.~\eqref{Eq shared energy bound} the required energy scales as $\frac{1}{4^{n+1}}$ while for independent couplings the energy scales as $\frac{1}{2^{n+1}}$ due to there being $2^{n+1}$ elements in the coefficient multiset.
Interestingly, we note that if the ancilla qubits are appropriately prepared (without error), and the system is not entangled to begin the process, the entanglement of the system remains zero throughout the evolution.  
However; if the ancilla qubits are imperfectly prepared, entanglement will be generated. This highlights how the realized entanglement is a result not only of the input states but also of the \emph{entangleability} of the control Hamiltonian.  \emph{Entangleability} describes a Hamiltonian's ability to induce entanglement in some system, whether or not entanglement is generated in a given process.
This notion is related to entangling power \cite{eisert2021entangling} and provides a framework for understanding the energetic advantage of non-local Hamiltonian structures.

\paragraph{Universal quantum computation}

Using a similar construction to the controlled-NOT gates, we can also explore the coefficient multisets for arbitrary controlled-Hadamard gates, $\text{C}^n\text{H}$, which are given by $\lambda_{C^n\text{H}} = \{2^n \cdot\frac{\pi}{2^{n+1}},2^n \cdot -\frac{\pi}{2^{n+1}} \}$.
Like the controlled-NOT gates, the coefficients required to implement the controlled-Hadamard gates decrease exponentially with the number of ancilla qubits utilized.
Finally, we observe that the arbitrarily controlled-Toffoli gates are connected to the controlled-NOT gates via the simple relation $\text{C}^n\text{Toff} = \text{C}^{n+2}\text{X}$.
The exponential decrease in the coefficients is immediately adopted. However, two of the previously prepared ancilla qubits are moved explicitly into the computational subspace of the overall Hilbert space.  

Having shown that, if arbitrarily large numbers of ancilla qubits are utilized, both the Toffoli and Hadamard gates can be implemented vanishingly low energy in the control field, and given that these gates form a universal gate set \cite{aharonov2003simple,shi2002both}, we can conclude that a universal quantum computer necessitates fundamentally no energy to control.  
At first, this statement is remarkable; we start by calculating the energy required to implement simple computations, revealing that they require energy, and by increasing the complexity of control scheme, we drive this energy towards zero.  
Effectively, we are offloading more of the control from our driving field and into the extrinsically controlled interactions of the qubits.

A natural thought is that we have simply moved the control energy from excitations in the classical electromagnetic field into the excited states of our ancilla qubits.
However, in principle, the qubits can be designed in such a way that the computational basis is energetically degenerate and in these cases there is no work associated in changing qubit states.
This result shows immediately that, for different control systems (either classical or quantum), the energetic characteristics of this driving system define the energy requirements for precise control.

We conclude this section by placing our results in the context of practical quantum computers.  
First, this treatment does not include the thermodynamic dissipation required to prepare or reset the qubits or other energy required to prepare the computing apparatus.
In addition, our result requires the precise implementation of $N$-qubit interactions, a requirement that is unlikely to be realized in the near future, although we note that uncontrolled interactions of this form are common in practice.
Lastly, these results assume sufficient suppression of the energy loss through dissipation channels and sufficient suppression of other noise sources.

\paragraph{Concluding remarks}

One of the key open questions in realizing useful quantum computers is how to assess the required resources to precisely control these systems. 
In this work, we have provided a method for determining the fundamental energy requirements for any given quantum computation, provided that we know the gate decomposition.
This is done by examining the coupling of a system of qubits to a classical electromagnetic field.
We have demonstrated that the \emph{entangleability} of the Hamiltonian implementing a quantum gate allows for more efficient computations and, via this construction, have demonstrated that simple gate sets are energetically unfavorable compared to monolithic gates.
By taking this analysis to the extreme, we have finally argued that a universal quantum computer has no fundamental energy requirements for control.

Beyond quantum computing, this work highlights how the structure of multi-partite Hamiltonians, namely their \emph{entangleability}, is a control resource.
This observation motivates investigations into the utility of \emph{entangleability} in optimal control for other quantum tasks. Since the control benefits are realized even if entanglement is not generated, we can also wonder if other scenarios where entanglement has been attributed as the underlying resource have a dual explanation in terms of the Hamiltonian structure of their generating processes.

\acknowledgments{We would like to thank E. Doucet for insightful discussions S.D. acknowledges support from the John Templeton Foundation under Grant No. 62422.}

\bibliography{references}

\newpage

\appendix
\paragraph{Deriving the control error} \label{Variance Derrivation}
In this End Matter, we derive Eq.~\eqref{Eq: ErrorTerm} by explicitly calculating $\text{var}(\lambda) = \braket{\lambda^2}- \braket{\lambda}^2$, where we have dropped the subscript over $i$ for simplicity, recalling that $a_k\ket{\alpha_k(t)}=\alpha_k\ket{\alpha_k(t)}$ we have,
\begin{equation}\label{eq: <lambda^2>}
    \begin{aligned}
        \braket{\lambda}^2 &= \left(\int_0^\tau dt \bra{\psi}   \sum_k  g_k (a_k e^{- i \omega_k t} + a_k^\dag e^{i \omega_k t}) \ket{\psi}  \right)^2\\
        &= \left(\int_0^\tau dt   \sum_k  g_k (\alpha_k e^{- i \omega_k t} + \alpha_k^* e^{i \omega_k t})   \right)^2\\
        &= \int_0^\tau \int_0^\tau dt dt'  \sum_{j,k}  g_k g_j \left(\alpha_k \alpha_j e^{- i (\omega_k t + w_j t') } \right.  \\
        & \quad \quad \quad \quad \quad \quad  \quad \left. + \alpha_k \alpha_j^* e^{- i (\omega_k t - w_j t')}  +  \text{h.c.} \right),
    \end{aligned}
\end{equation}
and,
\begin{equation} \label{eq: <lambda>^2}
    \begin{aligned}
    \braket{\lambda^2} &= \bra{\psi}\left(\int_0^\tau dt    \sum_k  g_k (a_k e^{- i \omega_k t} + a_k^\dag e^{i \omega_k t})   \right)^2\ket{\psi}\\
        &= \bra{\psi}\int_0^\tau \int_0^\tau dt dt'  \sum_{j,k}  g_k g_j \left(a_k a_j e^{- i (\omega_k t + w_j t') } \right. \\
        &\qquad \qquad \qquad \left.+  a_k a_j^\dag e^{- i (\omega_k t - w_j t')} + \text{h.c.} \right) \ket{\psi}\\
  &= \int_0^\tau \int_0^\tau dt dt'  \sum_{j,k}  g_k g_j \left(\left[\alpha_k \alpha_j e^{- i (\omega_k t + w_j t') } \right. \right.\\
  &\left.\left.+  \alpha_k \alpha_j^* e^{- i (\omega_k t - w_j t')} + \text{h.c.}\right]  +\, \delta_{j,k} e^{- i (-\omega_k t + w_j t') }   \right),
    \end{aligned}
\end{equation}
where in the last step we have utilized the fact that $[a,a^\dag] = 1$ to evaluate all raising and lowering operators explicitly.  
We immediately see that all terms in Eqs.~\eqref{eq: <lambda^2>} and \eqref{eq: <lambda>^2} are identical except for when $j = k$ which allows us to immediately compute their difference as,
\begin{equation}
\begin{aligned}
    \braket{\lambda^2}- \braket{\lambda}^2 &=  \sum_k \int_0^\tau\int_0^\tau dt dt' g_k^2 e^{i(\omega_k t' - \omega_k t)}\\
    &= \sum_k \left(\int_0^\tau dt g_k e^{-i\omega_k t}\right) \left(\int_0^\tau dt g_k e^{i\omega_k t}\right)\\
    &= \sum_k \left| \int_0^\tau dt g_k e^{- i \omega_k t}\right|^2.
    \end{aligned}
\end{equation}

\paragraph{Upper bound on the achieved gate error} \label{Loschmidt Bound}

Finally, we prove that the error defined in Eq.~\eqref{Error Definition} provides an upper bound of the realized gate error for any input state in the computational basis.
We begin by calculating,
\begin{equation} \label{eq appendix step1}
    \text{U}_{G^{'}} \text{U}_G^{\dag} =  \mathds{1} + \sum_{k=1}^\infty  \frac{1}{k!}\left(i\sum_i  \epsilon_i V_i \right)^k. \\
\end{equation}
By taking the expectation value of $\text{U}_{G^{'}} \text{U}_G^{\dag}$ with some normalized state in the computational basis $\ket{\psi}$ we can compute one minus this quantity as,
\begin{equation}
    \left|1  - \bra{\psi}\text{U}_{G^{'}} \text{U}_G^{\dag}\ket{\psi}\right| = \left|\bra{\psi}\sum_{k=1}^\infty  \frac{1}{k!} \left(i\sum_i  \epsilon_i V_i \right)^k \ket{\psi}\right|.
\end{equation}
Recalling the definition of the error \eqref{Error Definition} in combination with the definition of $\text{U}_{G^{'}} \text{U}_G^{\dag}$ from Eq.~\eqref{eq appendix step1},
\begin{equation}
    \epsilon = \left\Vert \sum_{k=1}^\infty  \frac{1}{k!}\left(i\sum_i \epsilon_iV_i \right)^k \right\Vert_\infty,
 \end{equation}
and noting that the operator norm identifies the largest possible change in norm on any non-zero state, i.e.  $||a||_{\infty} = \text{sup}\left(\frac{||A \ket{\psi}||}{||\ket{\psi}||} : \ket{\psi} \ne 0\right)$ we immediately have our desired result,
\begin{equation}
    \epsilon \ge \left|1  - \bra{\psi}\text{U}_{G^{'}} \text{U}_G^{\dag}\ket{\psi}\right|,
\end{equation}
for any state $\ket{\psi}$ in the computational basis.

\end{document}